\documentclass[aps,prd,onecolumn,groupedaddress,showpacs,nofootinbib,amssymb]{revtex4}
\usepackage[dvips]{graphicx}
\usepackage{amssymb}
\usepackage{amsmath}
\usepackage{graphicx,,color}
\usepackage{amsfonts}
\usepackage{bm}
\usepackage{cancel}
\usepackage{comment}

\newcommand\be{\begin{equation}}
\newcommand\ee{\end{equation}}

\newcommand\e{\mathrm{e}}

\allowdisplaybreaks[4]

\begin{document}

\tolerance=5000

\title{$k$-essence $f(R)$ Gravity Inflation}
\author{S.~Nojiri,$^{1,2}$\,\thanks{nojiri@gravity.phys.nagoya-u.ac.jp}
S.~D.~Odintsov,$^{3,4}$\,\thanks{odintsov@ieec.uab.es}
V.K.~Oikonomou,$^{5,6,7}$\,\thanks{v.k.oikonomou1979@gmail.com}}

\affiliation{$^{1)}$ Department of Physics, Nagoya University,
Nagoya 464-8602, Japan \\
$^{2)}$ Kobayashi-Maskawa Institute for the Origin of Particles
and
the Universe, Nagoya University, Nagoya 464-8602, Japan \\
$^{3)}$ ICREA, Passeig Luis Companys, 23, 08010 Barcelona, Spain\\
$^{4)}$ Institute of Space Sciences (IEEC-CSIC) C. Can Magrans
s/n,
08193 Barcelona, Spain\\
$^{5)}$ Department of Physics, Aristotle University of
Thessaloniki, Thessaloniki 54124,
Greece\\
$^{6)}$ Laboratory for Theoretical Cosmology, Tomsk State
University of Control Systems
and Radioelectronics, 634050 Tomsk, Russia (TUSUR)\\
$^{7)}$ Tomsk State Pedagogical University, 634061 Tomsk, Russia\\
}

\tolerance=5000

\begin{abstract}
In this work we study a modified version of $f(R)$ gravity in
which higher order kinetic terms of a scalar field are added in
the action of vacuum $f(R)$ gravity. This type of theory is a type
of $k$-essence $f(R)$ gravity, and it belongs to the general class
of $f(R,\phi ,X)$ theories of gravity, where $\phi$ is a scalar
field and $X=\frac{1}{2}\partial^{\mu}\phi\partial_{\mu}\phi$. We
focus on the inflationary phenomenology of the model, in the
slow-roll approximation, and we investigate whether viable
inflationary evolutions can be realized in the context of this
theory. We use two approaches, firstly by imposing the slow-roll
conditions and by using a non-viable vacuum $f(R)$ gravity. As we
demonstrate, the spectral index of the primordial scalar
perturbations and the tensor-to-scalar ratio of the resulting
theory can be compatible with the latest observational data. In
the second approach, we fix the functional form of the Hubble rate
as a function of the $e$-foldings number, and we modify well-known
vacuum $f(R)$ gravity reconstruction techniques, in order to find
the $k$-essence $f(R)$ gravity which can realize the given Hubble
rate. Accordingly, we calculate the slow-roll indices and the
corresponding observational indices, and we also provide general
formulas of these quantities in the slow-roll approximation. As we
demonstrate, viability can be obtained in this case too, however
the result is strongly model dependent. In addition, we discuss
when ghosts can occur in the theory, and we investigate under
which conditions ghosts can be avoided by using a particular class
of models. Finally, we qualitatively discuss the existence of
inflationary attractors for the non-slow-roll theory, and we
provide hints towards finding general de Sitter attractors for the
theory at hand.
\end{abstract}


\maketitle

\section{Introduction}

The primordial era of our Universe is one of the mysteries in
modern cosmology that need to be resolved. The recent
observational data \cite{Ade:2015lrj} have indicated that the
primordial curvature perturbations power spectrum is nearly scale
invariant, and there exist two kind of theories that can produce
such a nearly scale invariant power spectrum, the inflationary
theories \cite{Guth:1980zm,Starobinsky:1982ee,Linde:1983gd} and
bouncing cosmologies
\cite{Brandenberger:2016vhg,deHaro:2015wda,Cai:2014bea}. In
addition, both these candidate theories predict a small amount of
primordial gravitational radiation
\cite{Ade:2015lrj,Array:2015xqh}, so these two mainstream theories
could be viable candidate theories for the early-time Universe.
The inflationary theories have been studied for quite some time
\cite{Guth:1980zm,Starobinsky:1982ee,Linde:1983gd}, and there are
various gravitational theoretical frameworks which can produce an
early-time acceleration era, for example the modified gravity
framework
\cite{Nojiri:2017ncd,Nojiri:2010wj,Nojiri:2006ri,Capozziello:2011et,Capozziello:2010zz,delaCruzDombriz:2012xy,Olmo:2011uz}
and so on. With regard to the bounce cosmology alternative
description, these theories became popular after the Loop Quantum
Cosmology theory \cite{Ashtekar:2011ni,Ashtekar:2006wn,
Salo:2016dsr,Xiong:2007cn,Amoros:2014tha,Cai:2014zga,deHaro:2014kxa,
Kleidis:2018plu,Kleidis:2017ftt} resulted in the generation of a
quantum bounce. One of the theories that can also realize a
successful inflationary era, are the so-called $k$-essence
theories~\cite{Chiba:1999ka,ArmendarizPicon:2000dh,ArmendarizPicon:1999rj,ArmendarizPicon:2000ah,Chiba:2002mw,Malquarti:2003nn,Malquarti:2003hn,Chimento:2003zf,Chimento:2003ta,Scherrer:2004au,Aguirregabiria:2004te,ArmendarizPicon:2005nz,Abramo:2005be,Rendall:2005fv,Bruneton:2006gf,dePutter:2007ny,Babichev:2007dw,Deffayet:2011gz,Kan:2018odq},
which can also generate other appealing features of cosmological
evolution. In this paper we shall be interested in studying a
$k$-essence modified $f(R)$ gravity of a simple form, by adding a
higher order scalar field kinetic term, along with the vacuum
$f(R)$ gravity. The resulting theory is an $f(R,\phi,X)$ theory,
with $X=\frac{1}{2}\partial^{\mu}\phi\partial_{\mu}\phi$. The
cosmological perturbations of this kind of theories were derived
in
Refs.~\cite{Noh:2001ia,Hwang:2005hb,Hwang:2002fp,Kaiser:2013sna},
so our main aim in this paper is to investigate whether a viable
inflationary evolution can be realized in the context of the
$k$-essence $f(R)$ gravity. To this end, we investigate how the
slow-roll conditions modify the resulting equations of motion, and
we derive the solutions of the slow-roll theory, with regard to
the scalar field. After that we employ two different approaches in
order to study the phenomenological implications of the
$k$-essence $f(R)$ gravity theory. In the first approach, we
choose the functional form of the $f(R)$ gravity, and we
investigate how the $k$-essence term affects the cosmological
evolution in terms of the Hubble rate. After that we calculate in
detail the slow-roll indices of the inflationary theory at hand,
and correspondingly the observational indices. Eventually we
investigate the parameter space of the theory and we test the
phenomenological validity of the theory. The choice of the
functional form of the $f(R)$ gravity is such, so that the vacuum
$f(R)$ gravity is not phenomenologically viable, so in effect we
investigate whether the $k$-essence $f(R)$ gravity can be a
phenomenologically acceptable theory. In the second approach, we
fix the functional form of the Hubble rate as a function of the
$e$-foldings number, and we investigate which $k$-essence $f(R)$
gravity in the slow-roll approximation can produce such a
cosmological evolution. After this we express the slow-roll
indices as functions of the $e$-foldings number in the slow-roll
approximation, and we provide their functional form in detail, and
by using the resulting $f(R)$ gravity, we test the validity of the
theory by examining the parameter space. As we will demonstrate,
in this case too, it is possible to produce a viable inflationary
evolution in the context of $k$-essence $f(R)$ gravity. In
addition we examine the conditions under which ghosts can occur in
the theory, so we discriminate the ghost-free and phantom cases,
and the above considerations are given in terms of these two
cases.

This paper is organized as follows: In section II we investigate
when ghost degrees of freedom can occur in a general $k$-essence
$f(R)$ gravity, and we find the no-ghost constraints on a special
class of $k$-essence $f(R)$ gravity models. In section III we
present the essential features of the proposed $k$-essence $f(R)$
gravity theory, we derive the equations of motion and we
investigate how the slow-roll conditions affect the resulting
solution of the scalar field. After that we choose the functional
form of the $f(R)$ gravity and we calculate the slow-roll indices
of the resulting theory. Accordingly we calculate the
observational indices and we test the validity of the theory by
confronting it with the observational data. In section IV we use
another approach, by fixing the Hubble rate, and we investigate
which $k$-essence $f(R)$ gravity can produce such a cosmic
evolution. We provide detailed formulas for the slow-roll indices
as functions of the $e$-foldings number, and we calculate the
observational indices in the slow-roll approximation. Accordingly,
the viability of the theory is tested by confronting it with the
observational data. Finally, the conclusions follow in the end of
the paper.

Before we get to the core of this paper, we will discuss in brief
the geometric framework which shall be assumed in the rest of this
paper. We shall work with a flat Friedmann-Robertson-Walker (FRW)
metric, the line element of which is,
\begin{equation}
\label{metricfrw}
ds^2 = - dt^2 + a(t)^2 \sum_{i=1,2,3} \left(dx^i\right)^2\, ,
\end{equation}
with $a(t)$ being the scale factor as usual. Moreover, the metric
connection we will choose is the Levi-Civita, which is a
symmetric, metric compatible and torsion-less. Finally, the Ricci
scalar for the FRW metric of Eq.~(\ref{metricfrw}) is,
\begin{equation}
\label{ricciscalarform}
R=12 H^2+6\dot{H}\, ,
\end{equation}
where $H(t)$ is the Hubble rate, $H(t)\equiv \dot a(t)/ a(t)$ and
the ``dot'' indicates
differentiation with respect to the cosmic time.

\section{Ghosts in $k$-essence $f(R)$ Gravity and Conditions of Avoidance}

Before we start discussing the inflationary phenomenology of
$k$-essence $f(R)$ gravity models, we need to investigate when do
ghosts occur in the theory. In this section we shall discuss this
issue thoroughly for a class of $k$-essence $f(R)$ models.
Consider a general class of $k$-essence $f(R)$ gravity models of
the form,
\begin{equation}
\label{mainactionB} \mathcal{S}=\int d^4x\sqrt{-g}\left[
\frac{1}{2\kappa^2}f(R)+G(X) \right]\, ,
\end{equation}
where $\kappa^2=8\pi G$, $G$ is Newton's constant and also
$X=\frac{1}{2}\partial^{\mu}\phi\partial_{\mu}\phi$, with $\phi$
being a real scalar field. In order to investigate whether ghosts
can occur in this theory, we consider the perturbation of the
scalar field $\phi$ around the background solution $\phi=\phi_0$,
\begin{equation}
\label{per1} \phi = \phi_0 + \varphi \, .
\end{equation}
Then, due to the fact that,
\begin{equation}
\label{per2} X=\frac{1}{2}\partial^{\mu}\phi\partial_{\mu}\phi =
\frac{1}{2}\partial^{\mu}\phi_0 \partial_{\mu}\phi_0 +
\partial^{\mu}\phi\partial_{\mu}\varphi +
\frac{1}{2}\partial^{\mu}\varphi\partial_{\mu}\varphi \, ,
\end{equation}
we can expand the function $G(X)$ in the following way,
\begin{equation}
\label{per3} G(X) = G(X_0) + G_X (X_0)
\partial^{\mu}\phi_0\partial_{\mu}\varphi + \frac{1}{2} \left( G_X
(X_0) \partial^{\mu}\varphi\partial_{\mu}\varphi + G_{XX} (X_0)
\left( \partial^{\mu}\phi_0 \partial_{\mu}\varphi \right)^2
\right) + \mathcal{O} \left( \left( \varphi \right)^3 \right) \, ,
\end{equation}
where $X_0 \equiv \frac{1}{2} \partial^{\mu}\phi_0
\partial_{\mu}\phi_0$. The second term $G_X (X_0)
\partial^{\mu}\phi_0 \partial_{\mu}\varphi = \partial_\mu \left(
G_X (X_0) \partial^{\mu}\phi_0 \varphi \right)
 - \partial_\mu \left( G_X (X_0) \partial^{\mu}\phi_0 \right) \varphi$
becomes a total derivative due to the equation of motion
$0=\partial_\mu \left( G_X (X_0) \partial^{\mu}\phi_0 \right)$ so
the second term can be dropped. We may rewrite the third term as
follows,
\begin{equation}
\label{per4} \frac{1}{2} \left( G_X (X_0)
\partial^{\mu}\varphi\partial_{\mu}\varphi + G_{XX} (X_0) \left(
\partial^{\mu}\phi_0 \partial_{\mu}\varphi \right)^2 \right) =
\frac{1}{2} \mathcal{G}^{\mu\nu} \partial_\mu \varphi \partial_\nu
\varphi \, , \quad \mathcal{G}^{\mu\nu} \equiv G_X (X_0)
g^{\mu\nu} + G_{XX} (X_0) \partial^{\mu}\phi_0
\partial^{\nu}\phi_0 \, .
\end{equation}
Then in order to avoid having ghosts in the theory, we need to
require,
\begin{equation}
\label{per5} \mathcal{G}^{tt} = G_X (X_0) g^{tt} + G_{XX} (X_0)
\left( \dot\phi_0 \right)^2 > 0 \, .
\end{equation}
For the spatially flat FRW universe of Eq. (\ref{metricfrw}), if
we assume that $\phi_0$ depends solely on the cosmic time $t$, we
find,
\begin{equation}
\label{per5b} \mathcal{G}^{tt} = - G_X (X_0) + G_{XX} (X_0) \left(
\dot\phi_0 \right)^2\, , \quad X_0 = - \frac{1}{2} \left(
\dot\phi_0 \right)^2 <0 \, .
\end{equation}
In the following we shall consider models of the form,
\begin{equation}
\label{per6} G(X) = - X - \frac{1}{2}f_1X^m \, ,
\end{equation}
and also,
\begin{equation}
\label{per6a1} G(X) =  X + \frac{1}{2}f_1X^m \, .
\end{equation}
Obviously, the model of Eq. (\ref{per6a1}) contains a
non-canonical kinetic term for the scalar field, so the theory is
phantom from the beginning. However, the model of Eq. (\ref{per6})
can be ghost-free, so now we shall investigate the conditions
under which the theory is ghost-free. For the FRW background we
have,
\begin{equation}
\label{per7} \mathcal{G}^{tt} = 1 + \frac{m}{2} f_1 \left( -
\frac{1}{2} \left( \dot\phi_0 \right)^2 \right)^{m-1}
 + m(m-1) f_1 \left( - \frac{1}{2} \left( \dot\phi_0 \right)^2 \right)^{m-1}
= 1 + \left( m^2 + \frac{m}{2} \right) f_1 \left( - \frac{1}{2}
\left( \dot\phi_0 \right)^2 \right)^{m-1} \, .
\end{equation}
Therefore for the background solution $\phi=\phi_0$, if the
following condition holds true,
\begin{equation}
\label{per8} 1 + \left( m^2 + \frac{m}{2} \right) f_1 \left( -
\frac{1}{2} \left( \dot\phi_0 \right)^2 \right)^{m-1} > 0 \, ,
\end{equation}
no ghost occurs in the theory. In the next section, we shall prove
that the slow-roll solution for the model (\ref{per6}) and for $m$
even, has the form,
\begin{equation}
\label{solutionfinalslowrollscalarfieldsextranew} \phi (t)=\left(
2^{-m} m\,f_1 \right)^{\frac{1}{1-2 m}}\, t \, .
\end{equation}
For the slow-roll solution of Eq.
(\ref{solutionfinalslowrollscalarfields}), Eq.~(\ref{per8}) has
the following form,
\begin{equation}
\label{per8} 1 + \left( m^2 + \frac{m}{2} \right) \left(
f_1\right)^{- \frac{1}{1-2m}} 2^{\frac{1-m}{1 - 2m}}
m^{\frac{2(m-1)}{1-2m}} > 0 \, .
\end{equation}
Then if $m>0$ and even, and also for $f_1>0$, no ghost modes
appear in the theory.

Near stars or galaxies, $\phi_0$ might depend on the spatial
coordinates. In such a case, $X_0$ is not always negative but it
can be positive. Even in this case, if we assume (\ref{per6}), we
can write $\mathcal{G}^{tt}$ as follows,
\begin{equation}
\label{per10} \mathcal{G}^{tt} = - g^{tt} - g^{tt} \frac{m}{2} f_1
X_0^{m-1} - \frac{m(m-1)}{2} f_1 X_0^{m-2} \left( \dot\phi_0
\right)^2 = - g^{tt} + \frac{m}{2} f_0 X_0^{m-2} \left( g^{tt} X_0
- ( m-1) \left( \dot\phi_0 \right)^2 \right) \, .
\end{equation}
Then if,
\begin{equation}
\label{per11}
 - g^{tt} + \frac{m}{2} f_0 X_0^{m-2} \left( g^{tt} X_0 - ( m-1) \left( \dot\phi_0 \right)^2
\right) > 0 \, ,
\end{equation}
and no ghost occurs in this case too. Especially when $m$ is a
positive and even integer, if,
\begin{equation}
\label{per12} f_0>0\, , \quad g^{tt} X_0 - ( m-1) \left(
\dot\phi_0 \right)^2 > 0 \, ,
\end{equation}
or,
\begin{equation}
\label{per13} f_0<0\, , \quad g^{tt} X_0 - ( m-1) \left(
\dot\phi_0 \right)^2 < 0 \, ,
\end{equation}
no ghost occur in the theory. In summary, the case $m$ even and
positive and also if $f_1>0$ in the model (\ref{per6}) leads to a
ghost free theory, even at the astrophysical scales. We shall take
into account these constraints in the following sections.

\section{Slow-roll $k$-essence $f(R)$ Gravity: Model and Phenomenology}

The model of $k$-essence $f(R)$ gravity that we will study in this
work has the following action,
\begin{equation}
\label{mainaction} \mathcal{S}=\int d^4x\sqrt{-g}\left[
\frac{1}{2\kappa^2}f(R)\pm X\pm \frac{1}{2}f_1X^m\right] \, ,
\end{equation}
where $m$ is some positive number and the $\pm$ signs yield
different theories. As we demonstrated in the previous section,
the following model,
\begin{equation}
\label{mainactionghostfree} \mathcal{S}=\int d^4x\sqrt{-g}\left[
\frac{1}{2\kappa^2}f(R)- X- \frac{1}{2}f_1X^m\right] \, ,
\end{equation}
with $m$ an even integer, and $f_1>0$ leads to a ghost free
theory. Also we shall consider the phantom theory, in which case
the action is,
\begin{equation}
\label{mainactionphantom} \mathcal{S}=\int d^4x\sqrt{-g}\left[
\frac{1}{2\kappa^2}f(R)+ X+ \frac{1}{2}f_1X^m\right] \, ,
\end{equation}
with $f_1>0$, and in this section we shall investigate the
inflationary phenomenology of both the models
(\ref{mainactionghostfree}) and (\ref{mainactionphantom}). The
actions (\ref{mainaction}) belong to the general class of
$f(R,\phi,X)$ models of inflation, the cosmological perturbations
of which were extensively studied in
Refs.~\cite{Noh:2001ia,Hwang:2005hb,Hwang:2002fp,Kaiser:2013sna}.
In the following, we shall use the notation and formalism of
Refs.~\cite{Noh:2001ia,Hwang:2005hb,Hwang:2002fp,Kaiser:2013sna},
in order to study the phenomenology of the models
(\ref{mainactionghostfree}) and (\ref{mainactionphantom}). By
varying the action (\ref{mainactionB}) with respect to the metric,
also by using the FRW metric of Eq.~(\ref{metricfrw}), and finally
by assuming that the scalar field depends solely on the cosmic
time $t$, we obtain the following equations of motion,
\begin{align}
\label{equationsofmotion}
-\frac{1}{2}(f-F\,R)-\frac{\kappa^2}{2}G_{,X}(X)\dot{\phi}^2-3H\dot{F}=&3FH^2
\, , \nonumber \\
\ddot{F}-H\dot{F}+2\dot{H}F-\frac{\kappa^2}{2}G_{,X}(X)\dot{\phi}^2=& 0 \, ,
\nonumber \\
\frac{1}{a^3}\frac{d}{dt}(a^3G_{,X}(X)\dot{\phi})=&0\, ,
\end{align}
where $F(R)$, $G(X)$ and $G_{,X}(X)$ stand for,
\begin{equation}
\label{auxiliaryeqn} F=\frac{\partial f}{\partial R}\, , \quad
G(X)=\pm X\pm \frac{1}{2}f_1X^m\, ,\quad G_{,X}(X)=\frac{\partial
G}{\partial X}\, .
\end{equation}
Also, since the scalar field depends solely on the cosmic time,
the $k$-essence field $X$ is equal to $X=-\frac{1}{2}\dot{\phi}^2$.
We shall assume that the scalar field obeys the slow-roll
condition, which is,
\begin{equation}
\label{slowrollconditions}
\ddot{\phi}\ll H\dot{\phi}\, ,
\end{equation}
so let us see how the last equation in
Eq.~(\ref{equationsofmotion}) becomes in view of the condition
(\ref{slowrollconditions}). We shall discuss the implications of
the slow-roll condition (\ref{slowrollconditions}) on the
inflationary phenomenology for both the phantom theory
(\ref{mainactionphantom}) and for the ghost-free theory
(\ref{mainactionghostfree}).

\subsection{Ghost Free Inflation}

Let us study first the ghost-free slow-roll theory with action
(\ref{mainactionghostfree}), so let us rewrite it by using the
explicit form of the function $G(X)$, so in the case when $m$ is
an even integer, it reads,
\begin{align}
\label{thrirdequation} 0 = & 3 f_1 2^{-m} m H(t) \phi '(t)
\left(\phi '(t)^2\right)^{m-1}-3 H(t) \phi '(t)-\phi ''(t)  \\
\notag & f_1 2^{1-m} m^2 \phi '(t)^2- \phi ''(t)- \left(\phi
'(t)^2\right)^{m-2}-f_1 2^{1-m} m \phi '(t)^2 \phi ''(t)
\left(\phi '(t)^2\right)^{m-2}f_1 +2^{-m} m \phi ''(t) \left(\phi
'(t)^2\right)^{m-1}\, ,
\end{align}
So in view of the slow-roll condition (\ref{slowrollconditions}),
by dismissing terms containing the second derivative and higher
powers of the first derivative of the scalar field, we obtain,
\begin{equation}
\label{thridequationexplicit} 3 f_1 2^{-m} m H(t) \phi '(t)^{2
m}-3 H(t) \phi '(t)=0\, ,
\end{equation}
which can be solved and it yields,
\begin{equation}
\label{solutionslowroll} \dot{\phi}=\left(2^{-m} m\,f_1
\right)^{\frac{1}{1-2 m}}\, ,
\end{equation}
and by integrating with respect to the cosmic time we get the
solution,
\begin{equation}
\label{solutionfinalslowrollscalarfields} \phi (t)=\left( 2^{-m}
m\,f_1 \right)^{\frac{1}{1-2 m}}\, t \, .
\end{equation}
Hence the slow-roll condition for the theory at hand, uniquely
determines the evolution of the scalar field as a function of the
cosmic time. This will simplify significantly the calculation of
the slow-roll indices and of the corresponding observational
indices, as we show shortly.

\begin{figure}[h]
\centering
\includegraphics[width=18pc]{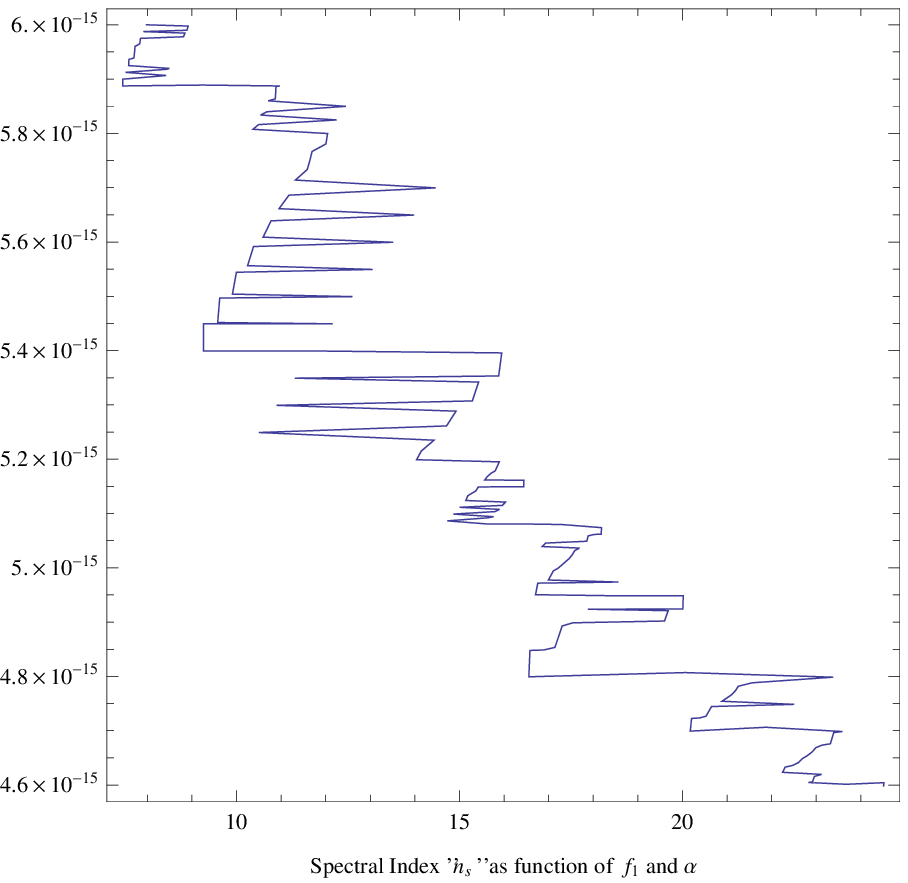}
\includegraphics[width=18pc]{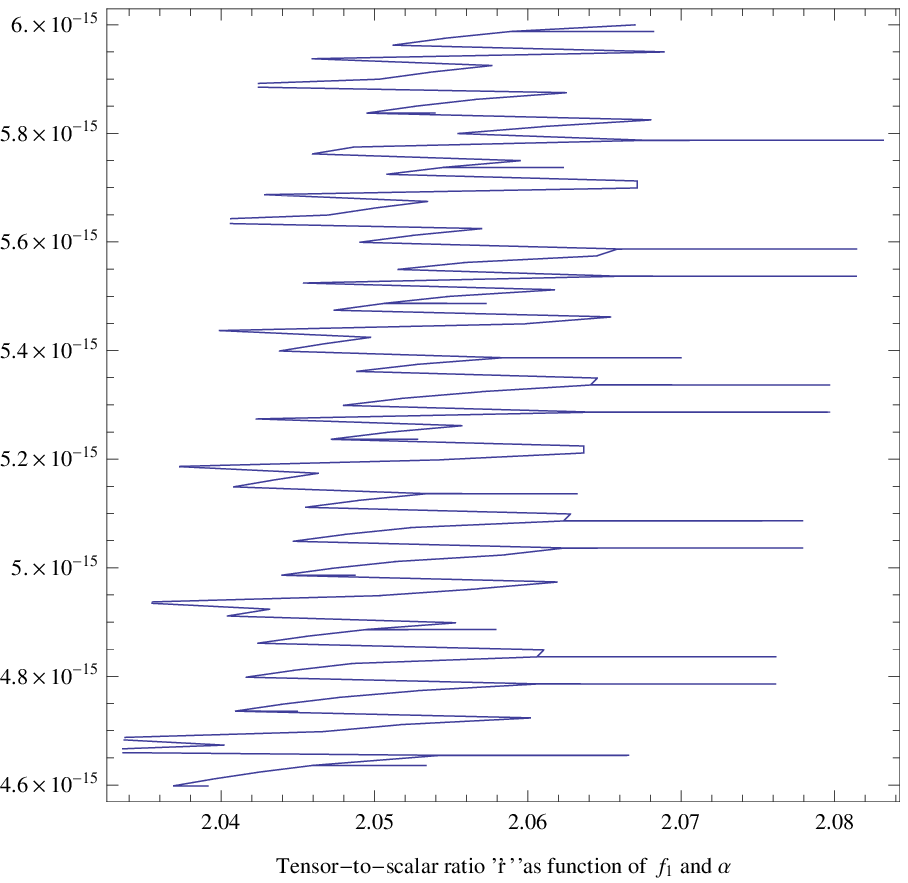}
\caption{The contour plot of the spectral index $n_s$ (left plot)
and of the tensor-to-scalar ratio (right plot) as functions of the
parameters $\alpha$ and $f_1$ for $\alpha=[4.59837\times
10^{-15},6\times 10^{-15}]$ and $f_1=[2.03291, 204]$, with
$(N,n,t_i,m)=(60,1.3660,10^{-25},2)$. } \label{newfigure1}
\end{figure}

Our aim is to investigate whether the addition of the $k$-essence
term $G(X)$ in a general $f(R)$ gravity, may eventually modify the
phenomenology of the vacuum $f(R)$ gravity. Thus, let us choose an
$f(R)$ gravity with problematic phenomenology, such as for example
the model,
\begin{equation}\label{frgravityexample}
f(R)=R+\alpha R^n\, ,
\end{equation}
where $\alpha,\, n>0$. Also in order to have inflation and not
superacceleration, the parameter $n$ is constrained to take values
in the interval $n= \left[\frac{1+\sqrt{3}}{2},2\right]$. The case
$n=2$ corresponds to the Starobinsky model
\cite{Starobinsky:1980te}, which gives a successful
phenomenological description for inflation, however the model
(\ref{frgravityexample}) has problematic inflationary
phenomenology for $\frac{1+\sqrt{3}}{2}\leq n <2$, due to the fact
one cannot obtain simultaneous overlap of the spectral index of
the primordial curvature perturbations and of the tensor-to-scalar
ratio with the Planck data, see \cite{Nojiri:2017ncd} for details
on this. Thus the main aim of this section is to show that the
$k$-essence modification of the $f(R)=R+\alpha R^n$ model may
alter its phenomenology. Let us start with the $k$-essence version
of the model $f(R)=R+\alpha R^n$, and the first equation of motion
of Eq.~(\ref{equationsofmotion}) in the slow-roll approximation
$\ddot{H}\ll H\dot{H}$, can be written as follows,
\begin{align}
\label{firsteqnaution$k$-essencefr}
0= & 3n \alpha R^{n-1}H^2 =\frac{\alpha (n-1)}{2}R^n-3n (n-1)\alpha
HR^{n-2}\dot{R} \nonumber \\
& -\frac{1}{2} \kappa ^2 \left(f_1 2^{-m} m\right)^{\frac{2}{1-2
m}} \left(f_1 2^{-m} m \left(f_1 2^{-m} m\right)^{\frac{2
(m-1)}{1-2 m}}-1\right)\, ,
\end{align}
where we used the explicit form of the scalar field $\phi (t)$ in
the slow-roll approximation, given in
Eq.~(\ref{solutionfinalslowrollscalarfields}). The last term in
Eq.~(\ref{firsteqnaution$k$-essencefr}) is subleading, therefore
the solution of Eq.~(\ref{firsteqnaution$k$-essencefr}) is the
following,
\begin{equation}
\label{solutionoffirsteqn}
H(t)=\frac{1}{c_1 \left(t-\frac{t_i}{c_1} \right)}\, ,
\end{equation}
where $t_i$ is some initial time and
$c_1=\frac{2-n}{(n-1)(2n-1)}$. The solutions
(\ref{solutionfinalslowrollscalarfields}) and
(\ref{solutionoffirsteqn}) will be very important for the
calculations of the slow-roll indices and of the corresponding
observational indices. Let us recall the functional form of the
slow-roll indices and of the corresponding observational indices,
for the theory with the action~(\ref{mainaction}). Following
Ref.~\cite{Noh:2001ia,Hwang:2005hb,Hwang:2002fp,Kaiser:2013sna},
the slow-roll indices $\epsilon_1$, $\epsilon_2$, $\epsilon_3$ and
$\epsilon_4$, are equal to,
\begin{equation}
\label{slowrollindicesmaineqn}
\epsilon_1=\frac{\dot{H}}{H^2}\, ,\quad
\epsilon_2=\frac{\ddot{\phi}}{H\dot{\phi}}\, , \quad
\epsilon_3=\frac{\dot{F}}{2HF}\, , \quad
\epsilon_4=\frac{\dot{E}}{2H E}\, ,
\end{equation}
where the function $E$ stands for,
\begin{equation}
\label{epsilonfunctione}
E=-\frac{F}{2X}\left(XG_{,X}+2X^2G_{,XX}+\frac{3\dot{F}^2}{2F}
\right)\, ,
\end{equation}
and we have set $\kappa^2=1$ for simplicity. Accordingly, the
spectral index of the primordial curvature perturbations $n_s$ and
tensor-to-scalar ratio $r$ are written in terms of the slow-roll
indices $\epsilon_1$, $\epsilon_2$, $\epsilon_3$ and $\epsilon_4$
as follows
\cite{Noh:2001ia,Hwang:2005hb,Hwang:2002fp,Kaiser:2013sna},
\begin{equation}
\label{observationalindices}
n_s=\frac{4\epsilon_1-2\epsilon_2+2\epsilon_3-2\epsilon_4}{1+\epsilon_1}\,
, \quad r=16 |\epsilon_1-\epsilon_3|c_A\, ,
\end{equation}
where $c_A$ stands for,
\begin{equation}
\label{cA}
c_A=\sqrt{\frac{XG_{,X}+\frac{3\dot{F}^2}{2F}}{XG_{,X}+2X^2G_{,XX}+\frac{3\dot{F}^2}{2F}}}
\, .
\end{equation}
By using the explicit form of the solutions
(\ref{solutionfinalslowrollscalarfields}) and
(\ref{solutionoffirsteqn}), the slow-roll indices
(\ref{slowrollindicesmaineqn}) expressed in terms of the
$e$-foldings number, take the following form,
\begin{align}
\label{slowrollindicesmaineqnexplicitform}
\epsilon_1=&-c_1\, ,\nonumber \\
\epsilon_2=&0\, , \nonumber \\
\epsilon_3=&-c_1 (n-1)\, , \nonumber \\
\epsilon_4=&\frac{2 (n-2) \left(2 J_1 (n-1) (1-2 n)^2+2 J_2 (n-1)
(1-2 n)^2+\alpha  12^n (n-2)^2 n (2 n-3) \left(\frac{\left(2 n^2-3
n+1\right)^2 e^{\frac{2 (n-2) N}{2 n^2-3 n+1}}}{\left(-2 n^2+2
n+1\right)^2 t_i^2}\right)^n\right)}{(n-1) (2 n-1) \left(2 J_1
(1-2 n)^2+2 J_2 (1-2 n)^2+\alpha  12^n (n-2)^2 n
\left(\frac{\left(2 n^2-3 n+1\right)^2 e^{\frac{2 (n-2) N}{2 n^2-3
n+1}}}{\left(-2 n^2+2 n+1\right)^2 t_i^2}\right)^n\right)}\, ,
\end{align}
where we introduced the parameters $J_1$, $J_2$, which are,
\begin{align}
\label{jparamtersng}
J_1=&\frac{1}{2} \left(f_1 2^{-m} m\right)^{\frac{2}{1-2 m}}-f_1 2^{-m-1} m \left(f_1 2^{-m} m\right)^{\frac{2 m}{1-2 m}}\, , \nonumber \\
J_2=&f_1 \left(-2^{-m}\right) (m-1) m \left(f_1 2^{-m} m\right)^{\frac{4}{1-2 m}} \left(f_1 2^{-m} m\right)^{\frac{2 (m-2)}{1-2 m}}\, , \nonumber \\
\, .
\end{align}
Having the slow-roll indices at hand, one can easily obtain the
observational indices (\ref{observationalindices}) in closed form,
however we do not quote these here since their final expressions
are too lengthy. The phenomenology of the resulting model is
interesting, due to the fact that by appropriately adjusting the
free parameters $f_1$, $\alpha$, $n$, $t_i$ and $m$, one can
obtain a viable phenomenology, having in mind the constraint on
the parameter $n$, which must take values in the range $n=
\left[\frac{1+\sqrt{3}}{2},2 \right]$. For example, by choosing
$n=1.36602$, $f_1=2.03291$, $\alpha=4.59843\times 10^{-15}$,
$t_i=10^{-25}$ and $m=2$, we obtain $n_s=0.965$ and $r=0.06$,
which are both compatible with the latest
Planck~\cite{Ade:2015lrj} and
BICEP2/Keck-Array~\cite{Array:2015xqh} data. In the ghost-free
model of (\ref{mainactionghostfree}) for the $f(R)\sim R^n$
gravity, the inflationary phenomenology is quite interesting and
the compatibility with the observational data comes easily without
any extreme fine-tuning of the free parameters. In fact, the
viability of the theory comes for a wide range of the free
parameters. This feature can be seen in Fig. \ref{newfigure1}
where we present the contour plots of the spectral index (left)
and of the tensor-to-scalar ratio (right) as functions of the
parameters $f_1$ and $\alpha$ for $f_1=[2.03291, 204]$ and
$\alpha=[4.59837\times 10^{-15},6\times 10^{-15}]$. Also the rest
of the parameters take the value $(N,n,t_i)=(60,1.3660,10^{-25})$.
The blue curves in the left plot correspond to the value
$n_s=0.965$ and the blue curves in the right plot correspond to
$r=0.06$.

\subsection{Phantom Inflation}

Now let us turn our focus on the phantom theory with action given
in Eq. (\ref{mainactionphantom}). For earlier works on $k$-essence
phantom theories see \cite{Aguirregabiria:2004te} and also Refs.
\cite{Liu:2010dh,Liu:2012iba,Piao:2004tq,Park:2018nfp} for general
phantom inflation models. We need to note that Also mention that
in principle, a phantom theory maybe just an effective description
and the complete theory may be free of ghosts.

In the phantom inflation case, if the slow-roll approximation is
assumed for the scalar field, the evolution of the phantom scalar
is governed by the following differential equation at leading
order,
\begin{equation}
\label{thridequationexplicitnewph} 3 H(t) \dot{\phi}(t)-3 f_1
2^{-m} m H(t) \dot{\phi}(t)^{2 m}=0\, ,
\end{equation}
which can be solved and it yields the same solution as in Eq.
(\ref{solutionfinalslowrollscalarfields}). Let us calculate the
slow-roll indices for the $f(R)=R+\alpha R^n$ model, so after
following the steps of the previous section, we obtain the
slow-roll indices (\ref{slowrollindicesmaineqnexplicitform}),
where in the case at hand, the parameters $J_1$, $J_2$ are,
\begin{align}
\label{jparamtersg} J_1=&f_1 2^{-m-1} m \left(\left(f_1
2^{-m} m\right)^{\frac{2}{1-2 m}}\right)^m-\frac{1}{2} \left(f_1 2^{-m} m\right)^{\frac{2}{1-2 m}}\, , \nonumber \\
J_2=&2^{-\frac{2 m^2}{1-2 m}-m} (m-1) m^{\frac{2 m}{1-2 m}+1}
f_1^{\frac{2 m}{1-2 m}+1}\, .
\end{align}
Accordingly one can easily obtain the observational indices
(\ref{observationalindices}) in closed form, which are too lengthy
to be presented here. By appropriately adjusting the free
parameters $f_1$, $\alpha$, $n$, $t_i$ and $m$, one can obtain a
viable phenomenology, for example, by choosing $n=1.36602$,
$f_1=10^{-40}$, $\alpha=6.751\times 10^{43}$, $t_i=10^{-20}$ and
$m=1.4$, we obtain $n_s=0.966$ and $r=0.0613$, which are both
compatible with the latest Planck~\cite{Ade:2015lrj} and
BICEP2/Keck-Array~\cite{Array:2015xqh} data. However, it is
obvious that extreme fine tuning is needed in the model,
nevertheless, a non-viable $f(R)$ gravity model becomes viable by
the inclusion of an appropriate phantom higher order kinetic
scalar field term in the gravitational action. In the next section
we shall present a general technique for obtaining viable
$k$-essence $f(R)$ gravity theories, in the slow-roll
approximation.

One issue we did address is the graceful exit from inflation issue
in the context of $k$-essence $f(R)$ gravity. Essentially from a
mathematical point of view, in order to have graceful exit from
inflation, the theory needs to have unstable de Sitter solutions
(which correspond to an effective equation of state parameter
$w_{eff}=-1$). This issue seems to depend strongly on the model of
$f(R)$ gravity chosen, and also depends on the slow-roll condition
and it's implications on the evolution of the scalar field at
early times. Formally, this problem can be answered in a concrete
way if one analyzes in detail the autonomous dynamical system of
the $k$-essence theory, find explicitly the de Sitter attractors
and investigate if these are stable or not. Also, the graceful
exit from inflation can be achieved by adding $R^2$ terms in the
gravitational action, however we do not discuss this issue further
in this paper and we hope to address in a more detailed future
work.

\section{An Alternative Approach to Slow-roll $k$-essence $f(R)$ Gravity Inflation}

\begin{figure}[h]
\centering
\includegraphics[width=18pc]{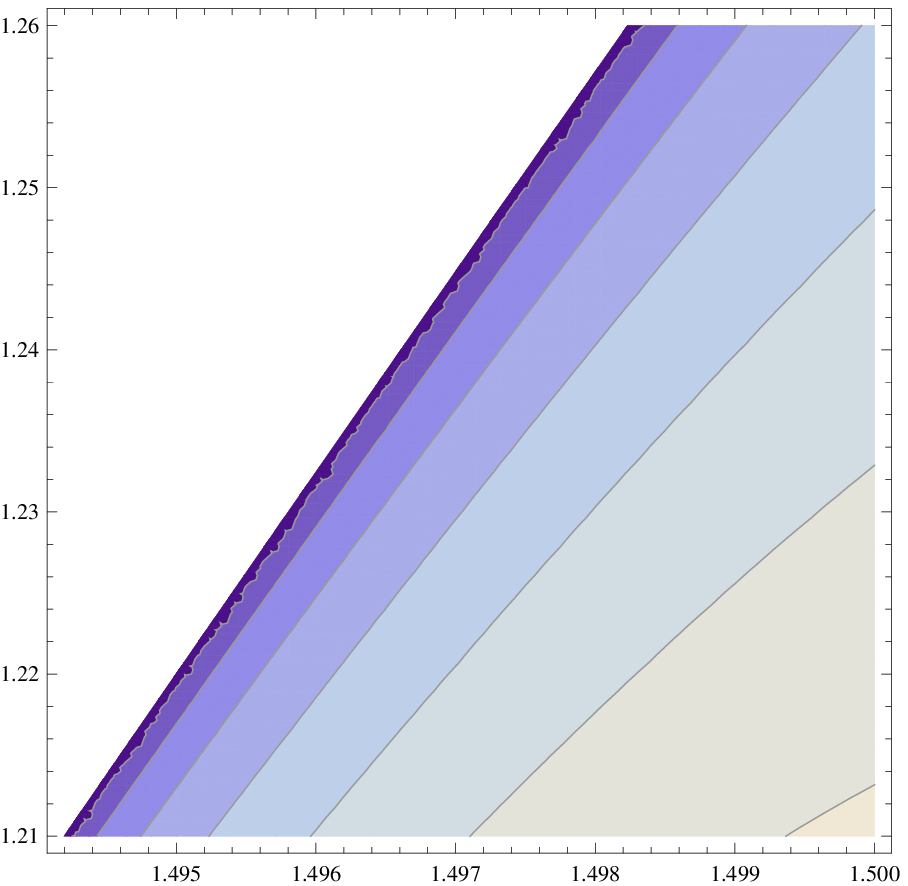}
\includegraphics[width=3.5pc]{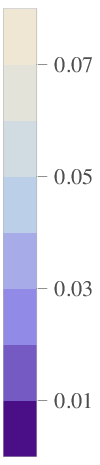}
\caption{The contour plot of the tensor-to-scalar ratio $r$ as a
function of the parameters $\beta$ and $f_1$ for $\beta=[1.49,
1.5]$, $f_1=[1.21, 1.26]$, with $(N,\gamma,m)=(60,0.001,1.2)$ and
$\mathcal{C}_1=\mathcal{C}_2=1$.}\label{plot2}
\end{figure}

In this section we shall employ a formalism appropriately designed
for the $k$-essence $f(R)$ gravity models of
Eqs.~(\ref{mainactionghostfree}) and (\ref{mainactionphantom}),
that will enable us to realize an arbitrarily given evolution and
also to test its viability. It is basically a reconstruction
technique for the $k$-essence $f(R)$ gravity theory (for general
reconstruction scheme for $k$-essence, see
\cite{Matsumoto:2010uv}), and we shall provide general formulas
that can be used for arbitrary forms of the $G(X)$ term. We start
off by providing the cosmological evolution we shall be interested
to realize with the theory at hand, which in terms of the
$e$-foldings number has the following form,
\begin{equation}
\label{hubblen}
H(N)=\gamma \e^{\frac{N}{4 \sqrt{3} \beta }}\, ,
\end{equation}
where $\beta$ and $\gamma$ are arbitrary parameters of the theory.
This cosmological evolution can be realized by specific
$k$-essence $f(R)$ gravities of the form
(\ref{mainactionghostfree}) and (\ref{mainactionphantom}) which we
will now find. To this end we shall appropriately modify the
reconstruction technique of Ref.~\cite{Nojiri:2009kx}, to
accommodate the $k$-essence term contribution, so we introduce the
function $G(N)=H(N)^2$, hence the Ricci scalar can be written as
follows,
\begin{equation}
\label{riccin}
R(N)=12G(N)+3G'(N)\, .
\end{equation}
Accordingly, by expressing the functions appearing in the first
equation of motion of Eq.~(\ref{equationsofmotion}), and also by
using the slow-roll solution of
Eq.~(\ref{solutionfinalslowrollscalarfields}) for the scalar
field, we obtain the following differential equation,
\begin{align}
\label{newfrw1modfrom} & -9G(N(R))\left ( 4G'(N(R))+G''(N(R))
\right )f''(R) +\left
(3G(N)+\frac{3}{2}G'(N(R)) \right )f'(R)\\
\notag & -\frac{f(R)}{2}+J_3^{\pm}=0\, ,
\end{align}
where $G'(N)=\mathrm{d}G(N)/\mathrm{d}N$ and
$G''(N)=\mathrm{d}^2G(N)/\mathrm{d}N^2$. Also $J_3^{\pm}$ in the
above equation is,
\begin{equation}\label{neweqnsq1}
J_3^{\pm}=2\,\kappa ^2 \left(2^{-\frac{2 m}{1-2 m}-1}\right)
m^{\frac{2}{1-2 m}} f_1^{\frac{2}{1-2 m}} \left(\pm 1-2^{-\frac{2
(m-1) m}{1-2 m}-m} m^{\frac{2 (m-1)}{1-2 m}+1} f_1^{\frac{2
(m-1)}{1-2 m}+1}\right)
\end{equation}
with the plus sign in the last term of Eq. (\ref{newfrw1modfrom}),
which is the $k$-essence term contribution, corresponding to the
phantom case (\ref{mainactionphantom}) while the minus sign
corresponding to the ghost-free theory
(\ref{mainactionghostfree}). Given the Hubble rate (\ref{hubblen})
and by inserting it in Eq.~(\ref{riccin}) we can find the function
$N(R)$ which reads,
\begin{equation}
\label{nr}
N(R)=2 \sqrt{3} \beta \ln \left(\frac{2 \beta R}{\left(24 \beta
+\sqrt{3}\right) \gamma ^2}\right)\, ,
\end{equation}
so accordingly, by using Eq.~(\ref{nr}) the differential equation
 (\ref{newfrw1modfrom}) becomes a second order differential
equation that can be solved to yield the exact $f(R)$ that can
realize the cosmological evolution (\ref{hubblen}). By combining
Eqs.~(\ref{hubblen}), (\ref{nr}) and (\ref{newfrw1modfrom}), we
get the following differential equation,
\begin{align}
\label{differenqlast} 0= & -\frac{\left(3 \left(8 \sqrt{3} \beta
+1\right) R^2\right) f''(R)}{\left(24 \beta
+\sqrt{3}\right)^2}+\frac{\left(\left(12 \beta +\sqrt{3}\right)
R\right) f'(R)}{2 \left(24 \beta +\sqrt{3}\right)}+J_3^{\pm}\, ,
\end{align}
which can be explicitly solved and it yields the solution,
\begin{equation}
\label{solutiondfiffeenq}
f(R)=\mathcal{C}_1R^{\mu}+\mathcal{C}_2R^{\nu}+\frac{2 \left(24
\beta  J_3^{\pm}+\sqrt{3} J_3^{\pm}\right)}{24 \beta +\sqrt{3}}\,
,
\end{equation}
where $\mathcal{C}_1$ and $\mathcal{C}_2$ are integration
constants, and also the parameters $\mu$ and $\nu$ appearing in
Eq.~(\ref{solutiondfiffeenq}) are defined as follows,
\begin{align}
\label{muandnu}
\mu=&\frac{96 \beta ^2+\frac{\sqrt{24 \beta +\sqrt{3}} \sqrt{384
\sqrt{3} \beta ^3-912 \beta ^2-32 \sqrt{3} \beta
+1}}{\sqrt[4]{3}}+28 \sqrt{3} \beta +3}{32 \sqrt{3} \beta +4}\, ,
\nonumber \\
\nu=& \frac{96 \beta ^2-\frac{\sqrt{24 \beta +\sqrt{3}} \sqrt{384
\sqrt{3} \beta ^3-912 \beta ^2-32 \sqrt{3} \beta
+1}}{\sqrt[4]{3}}+28 \sqrt{3} \beta +3}{32 \sqrt{3} \beta +4}\, .
\end{align}
Having the $f(R)$ gravity which realizes the cosmology
(\ref{hubblen}), we shall use the results of the slow-roll
formalism we developed in the previous section for the $k$-essence
$f(R)$ gravity, and we shall express the slow-roll indices and the
corresponding observational indices as functions of the
$e$-foldings number.
\begin{figure}[h]
\centering
\includegraphics[width=18pc]{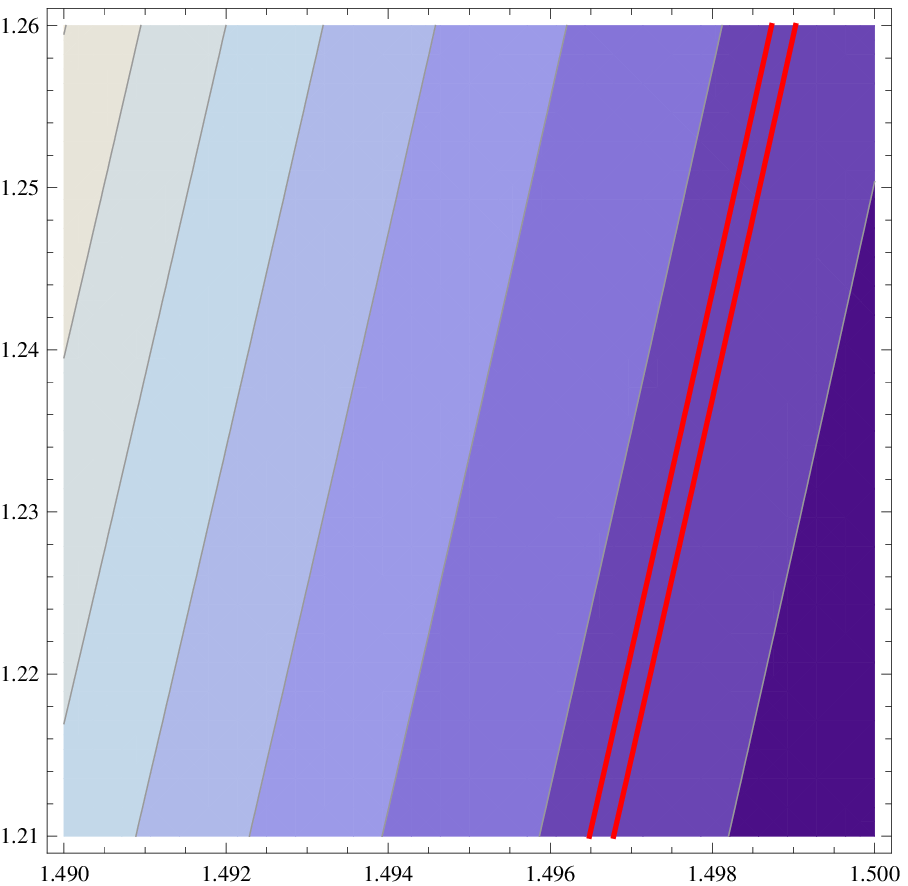}
\includegraphics[width=3pc]{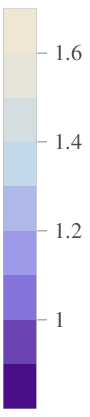}
\caption{The contour plot of the spectral index $n_s$ for the
phantom model (\ref{mainactionphantom}) as a function of the
parameters $\beta$ and $f_1$ for $\beta=[1.49, 1.5]$, $f_1=[1.21,
1.26]$, with $(N,\gamma,m)=(60,0.001,1.2)$ and
$\mathcal{C}_1=\mathcal{C}_2=1$. The red curves correspond to the
values of the spectral index $n_s=0.97161$ and $n_s=0.95839$,
which are the maximum and minimum values allowed by the Planck
data respectively.} \label{plot1}
\end{figure}
The formulas we shall produce will enable us to easily test the
viability of the resulting theory by confronting it with the
observational data of Planck~\cite{Ade:2015lrj}. By using the
following formula,
\begin{equation}
\label{trick1}
\frac{d}{dt}=H\frac{d}{dN}\, ,
\end{equation}
the slow-roll indices of Eq.~(\ref{slowrollindicesmaineqn}) can be
written in terms of the $e$-foldings number, and in the slow-roll
approximation these read,
\begin{align}
\label{slowrollindicesmaineqn123}
\epsilon_1=&\frac{H'(N)}{H(N)}\, ,\nonumber \\
\epsilon_2=&0\, , \nonumber \\
\epsilon_3=&\frac{12 H(N) H'(N) \left(\frac{d^2 F(R(N)}{d
R^2}\right)}{ F(R(N))}\, , \nonumber \\
\epsilon_4=&\frac{\frac{1728
H(N)^3 H'(N) \left(d^2 f(R(N))\right) \left(\frac{2 H'(N)^2
\left(d^2 f(R(N))\right)}{d R^2}+H(N) \left(\frac{H''(N) \left(d^2
f(R(N))\right)}{d R^2}+H(N) H'(N) \left(\frac{d^2 f(R(N))}{d
R^2}\right)'\right)\right)}{d R^2}+(J_1+J_2) F'(R(N))}{2 \left(864
H(N)^4 H'(N)^2 \left(\frac{d^2 f(R(N))}{d R^2}\right)^2+F(N)
(J_1+J_2)\right)}\, ,
\end{align}
where the prime indicates differentiation with respect to the
$e$-foldings number $N$, while the parameters $J_1$, $J_2$ are
defined in Eqs.~(\ref{jparamtersng}) and (\ref{jparamtersg}) for
the ghost-free and for the phantom case respectively. Accordingly,
the spectral index and the tensor-to-scalar ratio can be found by
the following formulas,
\begin{equation}
\label{nsandrintersmofn} n_s=\frac{2 (2
\epsilon_1+\epsilon_3-\epsilon_4)}{\epsilon_1+1}\, , \quad r=16
c_A |\epsilon_1-\epsilon_3|\, ,
\end{equation}
where $c_A$ is equal to,
\begin{equation}
\label{canewequationcrecon}
c_A=\sqrt{\frac{\frac{864 H(N)^4 H'(N)^2 \left(\frac{d^2
F(R(N))}{d R^2}\right)^2}{F(N)}+J_1}{\frac{864 H(N)^4 H'(N)^2
\left(\frac{d^2 F(R(N))}{d R^2}\right)^2}{F(N)}+J_1+J_2}}\, .
\end{equation}
For the case at hand, the Ricci scalar as a function of $N$ reads,
\begin{equation}
\label{riciasn}
R=12 \gamma ^2 \e^{\frac{N}{2 \sqrt{3} \beta }}+\frac{\sqrt{3}
\gamma ^2 \e^{\frac{N}{2 \sqrt{3} \beta }}}{2 \beta }\, ,
\end{equation}
therefore by using the explicit form of the $f(R)$ gravity
(\ref{solutiondfiffeenq}) and also by replacing $R(N)$ from
Eq.~(\ref{riciasn}), we can find the exact form of the slow-roll
indices and of the observational indices for both the ghost-free
theory (\ref{mainactionghostfree}) and for the phantom theory
(\ref{mainactionphantom}), which we do not quote here for brevity.
After a thorough investigation of the parameter space, it can be
seen that for the cosmological mode with Hubble rate
(\ref{hubblen}), both the phantom theory can provide a viable
phenomenology, in which a simultaneous compatibility of the
spectral index and of the tensor-to-scalar ratio with the
observational data can be achieved, for a wide range of
parameters. For example by using the following values for the free
parameters, $(N,\beta,\gamma,m,f_1)=(60,1.4983,0.001,1.2,1.2470)$
and by setting the integration constants
$\mathcal{C}_1=\mathcal{C}_2=1$, we obtain $n_s=0.964894$ and
$r=0.0179065$ which are compatible with the Planck data and also
with the BICEP2/Keck-Array data. This compatibility occurs for a
wide range of the free parameters, as it can be seen for example
in Figs.~\ref{plot1} and \ref{plot2}, where we present the contour
plots of the spectral index and of the tensor-to-scalar ratio for
$\beta$ chosen in the range $\beta=[1.49, 1.5]$ and for
$f_1=[1.21, 1.26]$ with $(N,\gamma,m)=(60,0.001,1.2)$ and
$\mathcal{C}_1=\mathcal{C}_2=1$. In Fig.~\ref{plot1}, the red
curves correspond to the values of the spectral index
$n_s=0.97161$ and $n_s=0.95839$, which are the maximum and minimum
values allowed by the Planck data respectively.
However, the ghost free theory does not provide simultaneous
compatibility of $n_s$ and $r$ with the observational data. For
example if one chooses,
$(N,\beta,\gamma,m,f_1,C_1,C_2)=(60,15.8,2,8,8,1,1)$, one obtains
$n_s=0.966282$ and $r=23$ which is an unappealing result. However
we need to note that the result is model dependent, so for the
specific cosmology which has the Hubble rate (\ref{hubblen}), it
seems that the phantom model (\ref{mainactionphantom}) provides
better phenomenology in comparison to the ghost-free model
(\ref{mainactionghostfree}).

Therefore, it is possible to produce viable inflationary
evolutions in the context of the $k$-essence $f(R)$ gravity, by
using the slow-roll formalism we presented in this section.
Basically, the method we presented is a reconstruction method for
realizing inflationary evolutions, in the slow-roll approximation.
In principle different types of inflationary evolutions can be
realized, but we refrain from going into details because the
procedure is the same as the example we presented.

\section{Conclusions}

In this paper we studied a modified gravity theoretical framework
which extends the vacuum $f(R)$ gravity theory, and it consists of
higher order scalar field kinetic terms that are added to the
standard $f(R)$ gravity Lagrangian density. Due to the form of the
extra terms in the action, we called this theory $k$-essence
$f(R)$ gravity theory, and our main aim was to investigate the
inflationary aspects of this theory, in the slow-roll
approximation. Actually, the class of $k$-essence $f(R)$ gravity
theory which we studied in this paper gets very much simplified if
the slow-roll condition is imposed on the scalar field, and we
investigated the dynamics of inflation in the resulting theory. By
using standard formulas for the slow-roll indices coming from
generalized $f(R,\phi,X)$ theories studied some time ago, we
derived the slow-roll indices for a general $f(R)$ gravity, and
then we applied the formalism for an $f(R)$ gravity of the form
$R+\alpha R^n$. This theory without the $k$-essence part is not
compatible with the latest Planck observational data, so we
questioned the viability of the theory in view of the presence of
the $k$-essence terms. As we demonstrated, there is a range of
values of the free parameters for which the phenomenological
viability of the theory can be achieved, for both the phantom and
ghost-free models which we used. Since the result might be model
dependent, we used another approach in order to see whether the
$k$-essence $f(R)$ gravity can produce viable phenomenology. To
this end, we fixed the Hubble rate as a function of the
$e$-foldings number, and we modified standard $f(R)$ gravity
reconstruction techniques to accommodate the presence of the
$k$-essence terms, always in the slow-roll approximation. Using
the resulting reconstruction techniques we derived the $k$-essence
$f(R)$ gravity which can realize the given Hubble rate, and then
we provided general formulas for the slow-roll indices as
functions of the $e$-foldings number, always in the slow-roll
approximation. Accordingly, we calculated the slow-roll indices
and the corresponding observational indices and we demonstrated
that the resulting theory can be compatible with the Planck data,
however the result is strongly model dependent. Thus we validated
that the $k$-essence $f(R)$ gravity theory can produce
phenomenologically viable cosmologies in the slow-roll
approximation. The latter is a vital ingredient of the formalism
we employed, so the basic question is, does this theory have
inflationary attractors in the absence of the slow-roll condition?
The vacuum $f(R)$ gravity theory has stable and unstable de Sitter
attractors without the slow-roll condition implied, as was
explicitly demonstrated in \cite{Odintsov:2017tbc}, by using the
dynamical system approach, so the question is does a general
non-slow-roll $k$-essence $f(R)$ gravity possesses inflationary
attractors? This question is non-trivial and no one can guarantee
this, before a consistent autonomous dynamical system is derived
for the theory in question. For example, in the case of
Gauss-Bonnet gravity there exist inflationary attractors even if
the slow-roll condition does not hold true, although these are
unstable, as was proved in Ref. \cite{Oikonomou:2017ppp}, and the
same applies for vacuum $f(R)$ gravity theories in the presence of
a non-flat metric \cite{workinprogress}. Moreover, the existence
of unstable de Sitter attractors is a feature of $f(R)$-$\phi$
theories \cite{Kleidis:2018cdx}. However the latter type of theory
contains potential terms, which are absent in the $k$-essence
$f(R)$ gravity, so the next major task is to question the
existence of inflationary attractors in the non-slow-roll
$k$-essence $f(R)$ gravity theory. To this end one should
appropriately construct a consistent autonomous dynamical system,
study its fixed points, and test their stability, analytically if
these are hyperbolic fixed points, or at least numerically if the
fixed points are non-hyperbolic. The interpretation of the
existence of unstable inflationary attractors is a major issue in
these theories, which in some sense can be viewed as an inherent
mechanism for the graceful exit from inflation, but this is a
highly non-trivial issue to discuss here, and of course out of the
context of this work. Work is in progress along the above research
lines.

Finally, it is noteworthy mentioning that even in this $k$-essence
framework, it is unavoidable having the initial Big Bang
singularity, when inflationary scenarios are considered. However,
it is interesting to note that, if the underlying theory can go
beyond the $k$-essence $f(R)$ gravity type inflation, namely a
torsional based $f(T)$ modified gravity
\cite{Cai:2011tc,Cai:2015emx}, or a Horndeski scalar
\cite{Cai:2012va} one may not only realize inflationary cosmology,
but also a non-singular bouncing phase that can be applied to
avoid the big bang singularity. In fact, it would be interesting
to extend the formalism we developed in this paper to find an
appropriate $k$-essence $f(R)$ gravity type theory that may
realize a bouncing cosmology. In the context of other extensions
of $f(R)$ gravity this is also possible \cite{Amoros:2014tha}, so
the question remains if there are $k$-essence modified gravities
that may realize cosmological bounces. We hope to address this
issue in a future work.

\section*{Acknowledgments}

This work is supported by MINECO (Spain), FIS2016-76363-P, and by
project 2017 SGR247 (AGAUR, Catalonia) (S.D.O). This work is also
supported by MEXT KAKENHI Grant-in-Aid for Scientific Research on
Innovative Areas ``Cosmic Acceleration'' No. 15H05890 (S.N.) and
the JSPS Grant-in-Aid for Scientific Research (C) No. 18K03615
(S.N.).

\end{document}